\documentclass[aps,twocolumn,english,superscriptaddress,citeautoscript,preprintnumbers,amsmath,amssymb,floatfix,footinbib]{revtex4-1}
\usepackage[breaklinks=true,colorlinks,citecolor=blue,linkcolor=blue,urlcolor=blue]{hyperref}
\usepackage{amsmath}
\usepackage{graphicx}
\usepackage{dcolumn}
\usepackage{float}
\usepackage[utf8]{inputenc}
\usepackage{xcolor}
\usepackage{soul}
\usepackage{ulem}
\usepackage{cancel}

\begin{document}
	
\title{ Weak antilocalization effect and triply degenerate state in Cu-doped CaAuAs}
\author{Sudip Malick}
\affiliation{Department of Physics, Indian Institute of Technology, Kanpur 208016, India}
\author{Arup Ghosh}
\affiliation{Saha Institute of Nuclear Physics, HBNI, 1/AF Bidhannagar, Calcutta 700 064, India}
\author{Chanchal K. Barman}
\affiliation{Department of Physics, Indian Institute of Technology, Bombay, Powai, Mumbai 400076, India}
\author{Aftab Alam}
\affiliation{Department of Physics, Indian Institute of Technology, Bombay, Powai, Mumbai 400076, India}
\author{Z. Hossain}
\email{zakir@iitk.ac.in}
\affiliation{Department of Physics, Indian Institute of Technology, Kanpur 208016, India}
\affiliation{Institute of Low Temperature and Structure Research,
	Polish Academy of Sciences, ulica Okolna 2, 50-422 Wroclaw, Poland}
\author{Prabhat Mandal}
\email{prabhat.mandal@saha.ac.in}
\affiliation{Saha Institute of Nuclear Physics, HBNI, 1/AF Bidhannagar, Calcutta 700 064, India}
\affiliation{Department of Condensed Matter Physics and Material Sciences, S. N. Bose National Centre for Basic Sciences, JD Block, Sector III, Salt Lake, Kolkata 700106, India}
\author{J. Nayak}
\email{jnayak@iitk.ac.in}
\affiliation{Department of Physics, Indian Institute of Technology, Kanpur 208016, India}

\begin{abstract}
The effect of 50\% Cu doping at the Au site in the topological Dirac semimetal CaAuAs is  investigated through electronic band structure calculations, electrical resistivity, and magnetotransport measurements. Electronic structure calculations a suggest broken-symmetry-driven topological phase transition from the Dirac to triple-point state in CaAuAs via alloy engineering. The electrical resistivity of both the CaAuAs and CaAu$_{0.5}$Cu$_{0.5}$As compounds shows metallic behavior. Nonsaturating quasilinear magnetoresistance (MR) behavior is observed in CaAuAs. On the other hand, MR of the doped compound shows a pronounced cusplike feature in the low-field regime. Such behavior of MR in CaAu$_{0.5}$Cu$_{0.5}$As is attributed to the weak antilocalization (WAL) effect. The WAL effect is analyzed using different theoretical models, including the semiclassical $\sim\sqrt{B}$ one which accounts for the three-dimensional WAL and modified Hikami-Larkin-Nagaoka model. Strong WAL effect is also observed in the longitudinal MR, which is well described by the generalized Altshuler-Aronov model. Our study suggests that the WAL effect originates from weak disorder and the spin-orbit coupled bulk state. Interestingly, we have also observed the signature of chiral anomaly in longitudinal MR, when both current and field are applied along the $c$ axis. The Hall resistivity measurements indicate that the charge conduction mechanism in these compounds is dominated by the holes with a concentration $\sim$10$^{20}$ cm$^{-3}$ and mobility $\sim 10^2$ cm$^2$ V$^{-1}$ S$^{-1}$.

\end{abstract}
	
	\maketitle
	
\section{INTRODUCTION}

Topological quantum materials have attracted considerable attention from the scientific community because they exhibit several novel electronic phenomena \cite{RevModPhys.82.3045, RevModPhys.90.015001}. The presence of distinct nontrivial electronic bands in these materials strongly influences transport behavior, resulting in extremely large linear and nonsaturating magnetoresistance, chiral anomalies, the topological Hall effect and the weak antilocalization effect (WAL) of the charge carrier \cite{Kumar2017,PhysRevB.96.241106,PhysRevB.96.075159,ZrSiS,TaAs,EuAgAs,CeAlGe,Mn3Sn,Cu-Sb2Te3,PhysRevB.87.035122}. These unconventional electronic bands introduce an additional nontrivial $\pi$ Berry phase around the time-reversal closed loop. In a weakly disordered system and quantum diffusion regime, the electronic conductivity is modified by the quantum interference of the carriers. This quantum interference leads to weak localization or weak antilocalization depending on the value of the phase shift of the carriers around the time-reversal loop. The nontrivial $\pi$ phase yields destructive interference, which is manifested through the WAL effect \cite{ma10070807,Hikami_1980,BERGMANN19841} in several three-dimensional (3D) topological insulators and semimetals \cite{Bi2Te3,BiSbTeSe2,PhysRevMaterials.4.034201,LaCuSb2,LuPtSb}.

Very recently, the WAL effect was realized in several ternary topological semimetals such as CaAgBi \cite{CaAgBi}, YbCdGe \cite{YbCdGe}, YbCdSn \cite{YbCdSn} and ScPdBi \cite{ScPdBi}. This particular ABC-type hexagonal family of materials provides various topological phases, including the Dirac semimetallic phase in CaAuAs, the topological insulating phase in CaAgAs, the nodal-line semimetallic phase in CaCdSn, and the correlated nodal-line semimetallic phase in YbCdGe \cite{AlB2, CaAuAs_Theo, CaAgAs, CaCdSn, YbCdGe}. Interestingly, various broken-symmetry-driven topological phases have also been observed in the ternary pnictide BaAgAs, and SrAgAs \cite{BaAgAs,SrAgAs}. Likewise, CaAuAs belonging to the $P6_3/mmc$ space group is an excellent example of a topological material that exhibits various topological ground states depending on the existing symmetry. According to a first-principles calculation, when the spin-orbit coupling (SOC) is ignored, CaAuAs hosts a nodal-line semimetallic state  \cite{CaAuAs_Theo}. Strikingly, a Dirac semimetallic state can also be achieved by considering SOC. Furthermore, in the absence of $C_3$ rotational symmetry and inversion symmetry, it leads to the topological insulating and Weyl semimetallic states, respectively  \cite{CaAuAs_Theo}. A recent report on angle-resolved photoemission spectroscopy (ARPES) corroborated the nontrivial band topology and Dirac semimetallic state in CaAuAs. However, the Dirac band crossing is observed about 0.25 eV above the Fermi level \cite{CaAuAs_ARPES}. It has been suggested that chemical doping in CaAuAs may shift the Fermi level, and the Dirac band crossing can be probed directly by studying ARPES and magnetotransport properties. Apart from the shift in the Fermi level, doping also introduces disorders in the crystal which may result in WAL or weak localization of the charge carrier. The above mentioned aspects encourage us to explore the effect of chemical doping on the magnetotransport properties of the topological Dirac semimetal CaAuAs.

In this work, we study the band structure and magnetotransport properties of a CaAuAs single crystal and examine the effect of chemical doping on CaAuAs by replacing 50\% of Au with Cu. The first-principles calculations reveal alloying-induced symmetry breaking in CaAuAs, and a phase transformation from the Dirac point to triple point (TP) occurs when 50\% Cu is doped at the Au site.  The magnetotransport measurements unfold strong a WAL effect up to room temperature in the doped compound. Furthermore,  a negative longitudinal MR is observed along the $c$ axis, which indicates a chiral anomaly effect. The Hall resistivity confirms the dominant hole type charge carrier with high mobility.

\section{SAMPLE PREPARATION AND EXPERIMENTAL TECHNIQUES}

\begin{figure}
	\includegraphics[width=8.6cm, keepaspectratio]{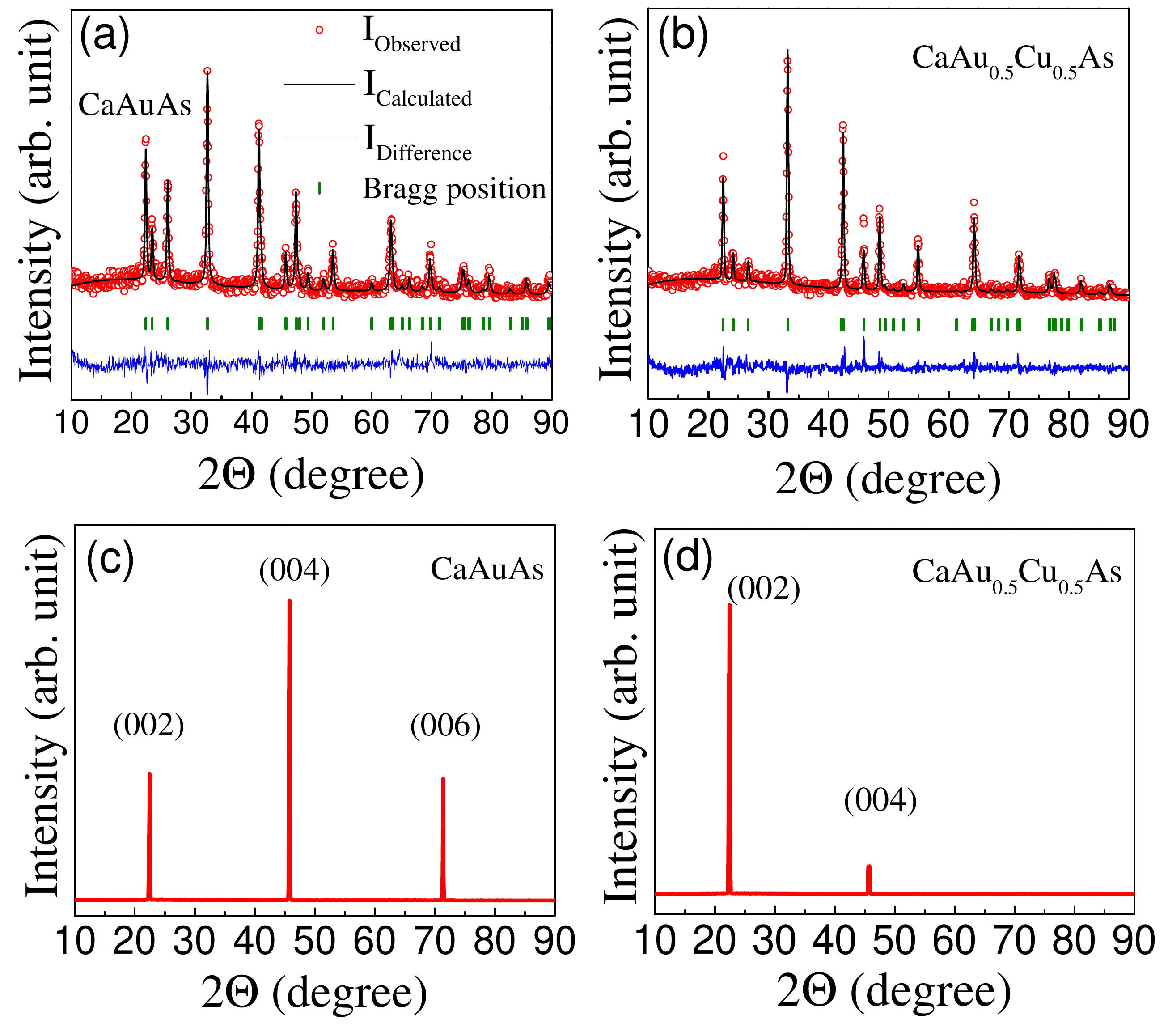}
	\caption{\label{XRD} (a) and (b) Rietveld refinement profile of the powder XRD pattern of crushed single crystals for CaAuAs and CaAu$_{0.5}$Cu$_{0.5}$As, respectively. Red dots are the experimental intensity, the black line is the calculated intensity, and the blue line is the difference between the experimental and calculated intensities. The green vertical lines display the Bragg positions. (c) and (d) The XRD pattern for CaAuAs and CaAu$_{0.5}$Cu$_{0.5}$As single crystals, respectively. Only (00$l$) peaks are observed.}
\end{figure}

Single crystals of CaAuAs and CaAu$_{0.5}$Cu$_{0.5}$As were grown by the flux method using bismuth as a flux. High-purity Ca (99.9\%, Alfa Aesar) shots, Au (99.999\%, Alfa Aesar) shots, Cu (99.9999\%, Alfa Aesar) shots, As (99.999\%, Alfa Aesar), and Bi (99.99\% Alfa Aesar) pieces were taken in 1:1:1:10, and 1:0.5:0.5:1:10 molar ratios for  CaAuAs and CaAu$_{0.5}$Cu$_{0.5}$As, respectively, and mixed thoroughly. The mixture was loaded into an alumina crucible and then sealed in an evacuated quartz ampule. The whole assembly was heated to 1050 $^\circ$C and soaked at this temperature for 20 h and then slowly cooled down to 400 $^\circ$C at a rate 3 $^\circ$C/h. Finally, the crystals were separated from the Bi flux by centrifuging. The typical size of the plate-like crystals is 1.5 mm$\times$1.5 mm$\times$0.2 mm. Further details of the crystal growth method can be found elsewhere \cite{CaAuAs_growth}. The grown crystals were characterized by x-ray diffraction (XRD) in a PANalytical X'Pert PRO diffractometer with Cu K$\alpha_1$ radiation and an energy dispersion x-ray spectroscopy (EDS) in a JEOL JSM-6010LA spectrometer. The electrical transport and magnetotransport measurements were carried out in a 9 T physical property measurement system (Quantum Design) using the standard four-probe technique using the ac-transport option. Electrical contacts were made using a thin gold wire and conducting silver paste. For the angle dependence of the MR between 0$^\circ$ and 360$^\circ$, a sample rotator was used.

The electronic band structure calculations were carried out using density functional theory implemented within the Vienna $ab$-initio simulation package (VASP) \cite{GKRESSE1993,JOUBERT1999}. The plane-wave basis set using the projected augmented wave \cite{PEBLOCH1994} method was used with an energy cut-off of 500 eV. Generalized-gradient approximation (GGA) \cite{JOUBERT1999} by Perdew, Burke, and Ernzerhof was employed to describe the exchange and correlations. The effect of spin-orbit coupling is explicitly included in all the calculations. The total energy (force) was converged up to 10$^{-5}$ eV (0.01 eV/\r{A}). The Brillouin zone (BZ) integrations were carried out using a $12\times 12\times 6$ $\Gamma$-centered $k$-mesh.

\section{RESULTS AND DISCUSSION}

\begin{table}
	\centering
	\caption {Lattice parameters $a$ and $c$ obtained from the refinement of the powder XRD data.}
	\label{table1}
	\vskip .2cm
	\addtolength{\tabcolsep}{+5pt}
	\begin{tabular}{c c c }
		\hline
		\hline
		& $a$ ({\AA})  &$c$ ({\AA}) \\[0.5ex]
		\hline
		CaAuAs                      & 4.3776(4)           & 7.939(3)  \\[1ex]
		CaAu$_{0.5}$Cu$_{0.5}$As    & 4.2587(4)	          & 7.912(3)  \\[1ex]
		
		\hline
	\end{tabular}
\end{table}

\subsection{Crystal Structure}

The powder x-ray diffraction pattern of crushed single crystals of CaAuAs and CaAu$_{0.5}$Cu$_{0.5}$ are shown in Figs. \ref{XRD} (a) and \ref{XRD} (b), respectively. Within the resolution of XRD, we did not see any peak due to the impurity phase. The diffraction pattern was analyzed by the Rietveld refinement method using FullProf software, which indicates that both the samples crystallize in a hexagonal structure with space group $P6_3/mmc$ (No. 194), and the corresponding lattice parameters are listed in Table \ref{table1}. The lattice parameters of pure compound agree well with the previous report \cite{AlB2}. The lattice constants $a$ and $c$ are reduced by 2.72\%, and 0.34\% ,respectively, after Cu doping in CaAuAs, which is expected as Cu has a smaller atomic radius than Au. In order to check the crystalline quality, we performed XRD on single crystals [Figs. \ref{XRD} (c) and (d)]. It is clear from the Fig. \ref{XRD}  that only (00$l$) peaks are present and they are extremely sharp. This confirms high crystalline quality of the grown single crystals. For the detailed information about the overall chemical composition, we determined the elemental concentrations at different randomly selected regions of the grown crystals using EDS. The EDS data confirm that the atomic compositions of CaAuAs and CaAu$_{0.5}$Cu$_{0.5}$As are close to the expected stoichiometries of 1:1:1 and 1:0.5:0.5:1, respectively.

\subsection{Electronic structure}

\begin{figure}
	\centering
	\includegraphics[width=0.45\textwidth]{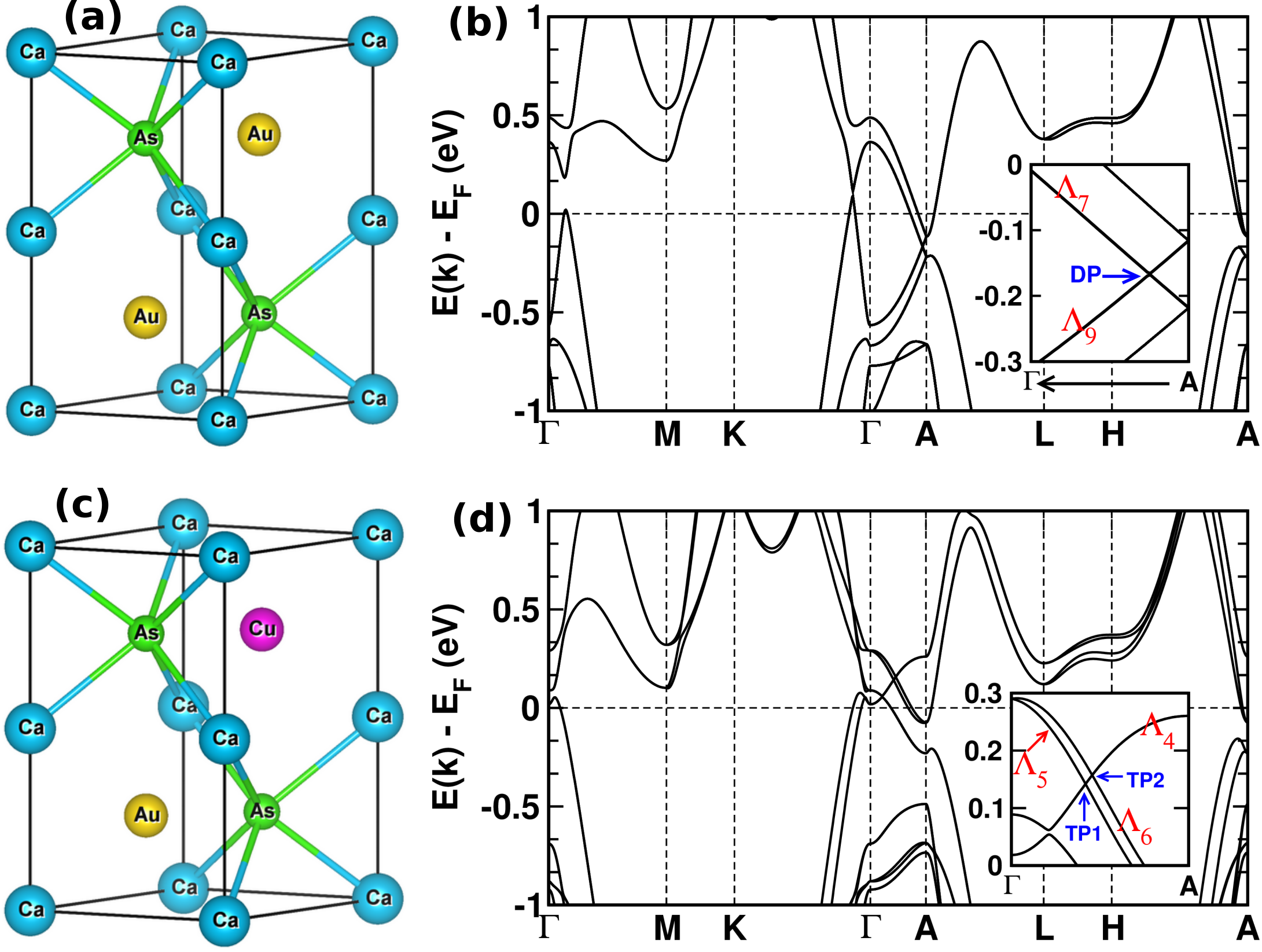}
	\caption{Crystal structure of (a) CaAuAs and (c) ordered  CaAu$_{0.5}$Cu$_{0.5}$As. (b) and (d) show the electronic band structure of CaAuAs and CaAu$_{0.5}$Cu$_{0.5}$As respectively, including the SOC effect. The insets in (b,d) show a zoomed-in view of the bands along $\Gamma$-A. $\Lambda_i,s$ are the irreducible representations of band characters. The Dirac point (DP) and triply degenerate nodal points (TP) are shown by arrows.}
	\label{Bands}
\end{figure}

Although XRD suggests that both the pure and doped compounds form in the same crystal structure, for electronic structure calculations, we have fully relaxed the lattice geometry to achieve the minimum energy structure. The optimized lattice parameters of CaAuAs are found to be $a=b=4.45$ \r{A} and $c=8.04$ \r{A}. The parent compound CaAuAs has 6 atoms in its primitive cell. To simulate CaAu$_{0.5}$Cu$_{0.5}$As, we have replaced one of the symmetry equivalent Au sites in the parent compound with a Cu atom. The optimized lattice parameters of the Cu-substituted compound are found to be $a=b=4.35$ \r{A} and $c=8.00$ \r{A}. These lattice parameters match the experimental data fairly well. Due to the underbinding effect of GGA, the theoretical lattice constants are slightly overestimated as compared to the experimental values. Figures. ~\ref{Bands} (a) and (c) show the crystal structure for the compounds.

{\par} The parent compound CaAuAs belongs to the $D_{6h}$ point group symmetry, which involves space inversion symmetry, and six fold rotational symmetry ($C_{6z}$) with respect to the \emph{z}-axis. The time reversal symmetry and space inversion symmetry together ensure Kramer's double degeneracy throughout the BZ of CaAuAs. The stable intercrossing of two doubly degenerate bands in the BZ is expected to give a four-fold degenerate Dirac point (DP). Along $\Gamma$-A, the site symmetry group is $C_{6v}$ and every band along this direction can be differentiated by the irreducible representations (IRREPS) of $C_{6v}$ symmetry. Fig.~\ref{Bands}(b) shows the electronic band structure for CaAuAs, including SOC. Clearly, the two-dimensional IRREPS $\Lambda_7$ and $\Lambda_9$ along the $\Gamma$-A direction intercross each other to form a four-fold degenerate DP. The stability of the DP is ensured by the group orthogonality which states that the hybridization between two distinct IRREPS is forbidden \cite{dresselhaus2007group}.

{\par} Insertion of 50\% Cu in CaAuAs (to form the alloy CaAu$_{0.5}$Cu$_{0.5}$As) breaks the space inversion symmetry. The alloy now hosts three-fold ($C_{3z}$) rotational symmetry and $C_{3v}$ site symmetry group along the \emph{z}-axis ($\Gamma$-A direction). Since the space-inversion symmetry is absent, the Kramer's degeneracy is lifted at generic momenta except time-reversal invariant points. However, the presence of $C_{3v}$ point group symmetry along the $\Gamma$-A direction allows both non-degenerate and doubly degenerate bands along this direction in the BZ. Fig.~\ref{Bands}(d) shows the electronic band structure of CaAu$_{0.5}$Cu$_{0.5}$As. The inset shows a zoomed-in view of the bands along the $\Gamma$-A direction. It clearly shows a doubly degenerate $\Lambda_4$ and two non-degenerate $\Lambda_5$ and $\Lambda_6$ bands. The intercrossing of the later two with $\Lambda_4$ band gives two triply degenerate nodal points TP1 and TP2. Using the symmetry analysis, it can be shown that the TPs are protected by the $C_{3v}$ point group symmetry of CaAu$_{0.5}$Cu$_{0.5}$As.

\subsection{Transport properties}

\begin{figure}
	\includegraphics[width=8cm, keepaspectratio]{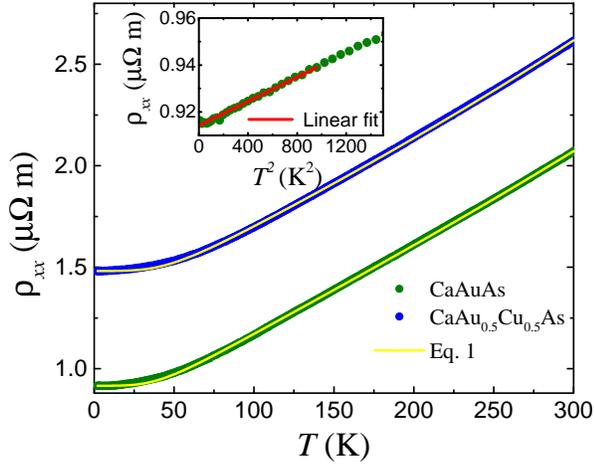}
	\caption{\label{RT} The electrical resistivities from 2 to 300 K for CaAuAs and the doped sample are shown. The yellow lines are the fitting with the BG expression plus an $aT^2$ term as given in Eq. \ref{Eq1}. $\rho_{xx}$ as a function of $T^2$ of CaAuAs is presented in the inset. The linear fit suggests quadratic behavior of $\rho_{xx}$ with  temperature ($T$ $<$ 30 K).}
\end{figure}

The temperature dependence of the zero-field electrical resistivity ($\rho_{xx}$) for CaAuAs and CaAu$_{0.5}$Cu$_{0.5}$As single crystals in the range 2 - 300 K is shown in Fig. \ref{RT}. With a decrease in temperature, $\rho_{xx}$ for both the samples deceases monotonically down to the lowest measured temperature, indicating metallic behavior.  It is clear from the figure that the doped compound has a larger residual resistivity. This is related to the disorder effect induced by Cu-doping. We have addressed the disorder effect in detail while discussing the magnetotransport property. The resistivities of both compounds follow a similar temperature dependence. The overall behavior of resistivity is quite similar to that observed in the topological materials CaAgBi \cite{CaAgBi} and CaAgAs \cite{CaAgAs}. $\rho_{xx}(T)$ is approximately linear at high temperatures but a weak upward curvature appears in the low-temperature region. We observe that the resistivity below 30 K exhibits $T^2$ dependence (see the inset in Fig. \ref{RT}). This behavior indicates a crossover from electron-phonon to electron-electron scattering mechanism with the decrease in temperature.  Thus, the resistivities for both compounds can be described by the Bloch Gr\"uneisen (BG)  scattering model along with a $T^2$ term, which is as follows,
\begin{equation}
\label{Eq1}
\rho_{xx} (T)= \rho_0 + 4R\left( \frac{T}{\Theta_R}\right)^5J_5\left( \frac{\Theta_R}{T}\right) +aT^2,
\end{equation}

where $J_5\left( \frac{\Theta_R}{T}\right)$ is the Gr\"uneisen integral function, $\Theta_R$ is the Debye temperature, and $R$ and $a$ are the constants. The first term $\rho_0$  of the above equation is the residual resistivity, second term represents the resistivity due to electron-phonon scattering and the third term is the contribution to the resistivity due to electron-electron scattering. The obtained fitting parameters from the Eq. \ref{Eq1} are $\rho_0$ = 0.916 $\mu\Omega$ m, $R$ = 0.83 $\mu\Omega$ m, $\Theta_R$ = 255 K, and $a$ = 4.40 $\times$10$^{-6}$ $\mu\Omega$ m K$^{-2}$, for CaAuAs and $\rho_0$ = 1.481 $\mu\Omega$ m, $R$ = 0.86  $\mu\Omega$ m, $\Theta_R$ = 294 K, and $a$ = 3.42$\times$10$^{-6}$ $\mu\Omega$ m K$^{-2}$ for CaAu$_{0.5}$Cu$_{0.5}$As.

\begin{figure}
	\includegraphics[width=7cm, keepaspectratio]{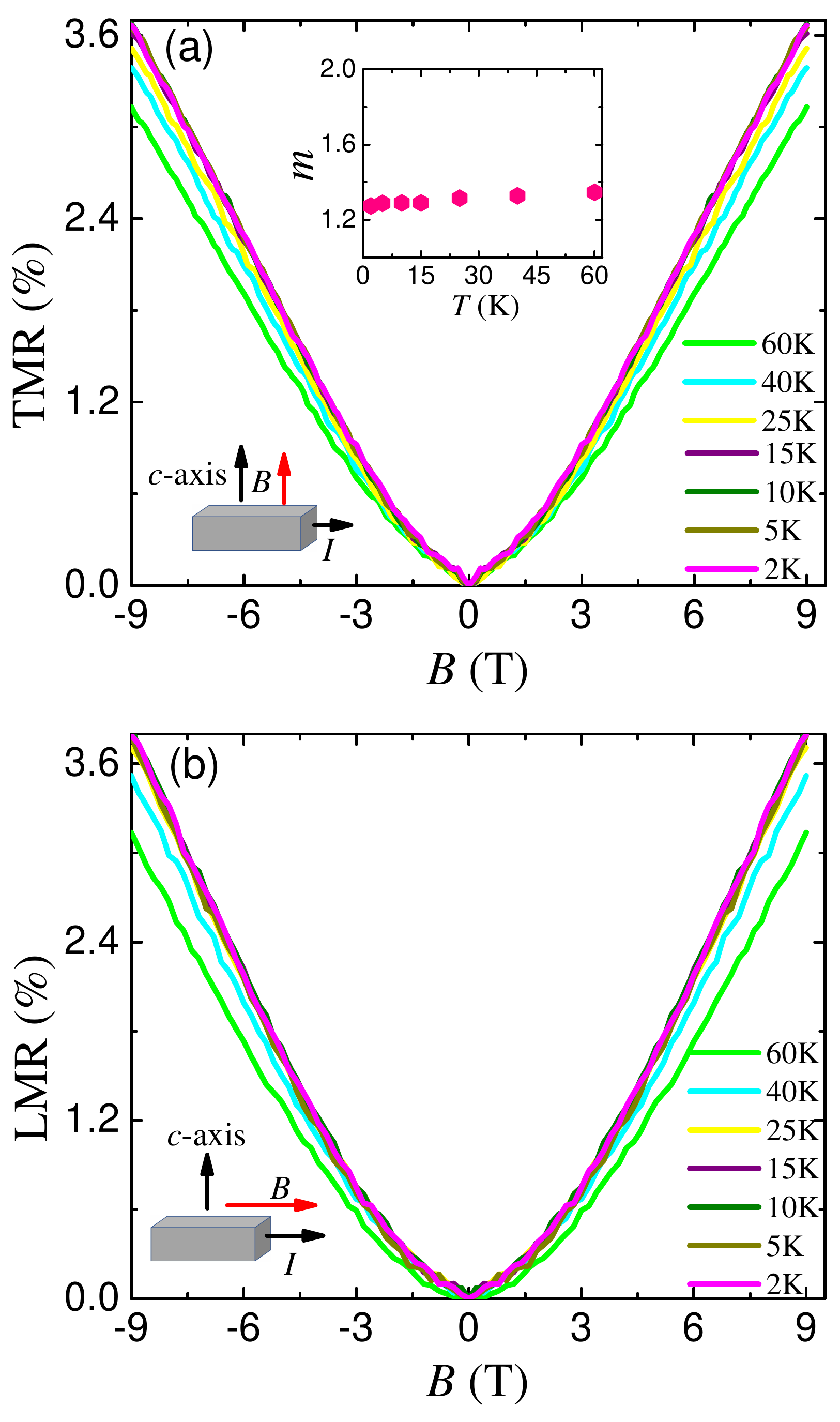}
	\caption{\label{MR_CaAuAs} The magnetic field dependence of TMR ($B\perp I$) and  LMR ($B\parallel I$) of CaAuAs crystal at different fixed temperatures in the range 2 - 60 K. The top inset in (a) shows the variation of the exponent $m$ with temperature. The measurement configurations for TMR and LMR are shown schematically in their respective insets.}
\end{figure}

\begin{figure*}
	\centering
	\includegraphics[width=15cm, keepaspectratio]{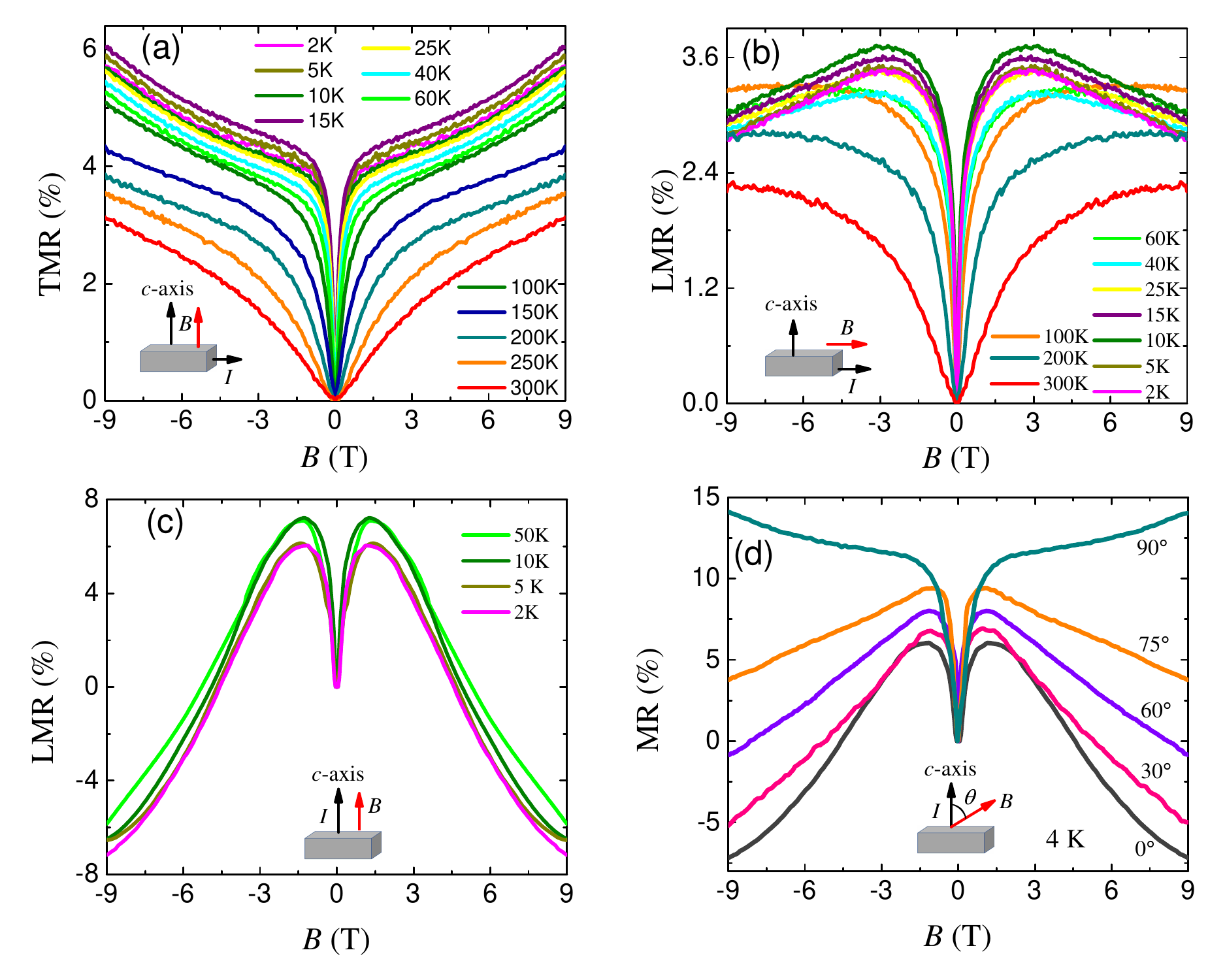}
	\caption{\label{MR_doped} The magnetic field dependence of TMR (a) and LMR (b) of CaAu$_{0.5}$Cu$_{0.5}$As crystal in the magnetic field range from -9 T to +9 T at few representative temperatures in the range 2 K - 300 K. (c) The field-dependent LMR along the $c$ axis measured in-between -9 T to +9 T at few fixed temperatures. (d) Field dependence of MR at 4 K for various field configurations when current is applied along the \textit{c} axis.}
\end{figure*}

Fig. \ref{MR_CaAuAs} displays the magnetic field ($B$) dependence of the transverse magnetoresistance (TMR) and longitudinal magnetoresistance (LMR) of CaAuAs at different temperatures. The directions of $B$ and the current ($I$) are schematically shown in the insets. MR is calculated using the expression MR = $[\rho_{xx}(B)-\rho_{xx}(0)]/\rho(0)$, where $\rho_{xx}(B)$ and $\rho_{xx}(0)$ are the resistivities in the presence and absence of the magnetic field, respectively. To eliminate the Hall resistivity contribution in MR, all the MR data were symmetrized using the formula $\rho_{xx}(B)$ = [$\rho_{xx}(+B)+\rho_{xx}(-B)$]/2.  Both TMR and LMR show a weak dependence on the magnetic field as well as the temperature. The maximum value of MR lies between 3\% and 4\% at 9 T. A similar low-value of MR is also found in other compounds such as CaIr$_2$Ge$_2$, ScPtBi, and LaAlSi \cite{CaIr2Ge2,ScPtBi,LaAlSi}. The observed value of MR is small compared to that reported in many topological semimetals. Extremely large MR is commonly found in topological materials mainly due to the perfect electron-hole compensation or very large carrier mobility \cite{PtBi2,WTe2,LaSb,YBi,TaP2,NbAs2}. CaAuAs excludes these facts as the dominant hole type of carrier is present, which has been confirmed from the Hall resistivity measurements, which will be discussed later. The low carrier mobility ($\mu$) is the dominating factor for observing low MR in both compounds as magnetoconductivity $\sigma(B) \sim n\mu/(1+\mu^2B^2)$ for a single-band model with carrier density $n$ \cite{jacoboni2010theory}. Furthermore, the fact that the Dirac band is about 250 meV above the Fermi level, might also be responsible for the low value of MR as the topological features may have little influence on the magnetotransport properties  \cite{CaAuAs_ARPES}. However, MR continuously increases with the applied field without any sign of saturation. To know the exact dependence of TMR on applied magnetic field, we have fitted the MR data using the expression $MR\propto B^m$ in the high field region (5 T $\leq B \leq$ 9 T). The obtained values of $m$ are in between 1.22 and 1.25 in the temperature range of 2  to 60 K, which is shown in the inset in Fig. \ref{MR_CaAuAs}(a). Such a nonsaturating quasi-linear behavior of MR is quite common in topological materials \cite{Cd3As2, CaCdSn}. It is likely that the quasi-linear behavior of MR in CaAuAs is due the to non-trivial topological state \cite{CaAuAs_ARPES,CaCdSn}.

We now focus on the magnetotransport properties of the doped sample, CaAu$_{0.5}$Cu$_{0.5}$As. Interestingly, the nature of the field dependence of MR after 50\% Cu-doping in CaAuAs has changed drastically. The magnetic field variation of TMR and LMR inthe range of $\pm$9 T are presented in Figs. \ref{MR_doped} (a) and (b). In the low-field region and below $\sim$100 K, both TMR and LMR increase very sharply with the increase of the magnetic field, and, as a result, a cusp-like feature appears. Such a cusplike feature in MR at low fields is attributed to the WAL effect. The WAL phenomenon was recently realized in several topological insulators and semimetals. However, very few systems show the WAL effect up to room temperature as we observe in CaAu$_{0.5}$Cu$_{0.5}$As \cite{CaAgBi, ScPtBi}. Moreover, TMR increases gradually as the magnetic field increases without saturating, reaching $\sim$ 6\% at 2 K in 9 T. The maximum value of TMR decreases as the temperature rises, eventually falling to $\sim$ 3\% at 300 K. Surprisingly, LMR shows a downturn at low temperatures for applied field above 3 T [Fig. \ref{MR_doped} (b)]. Such  behavior in LMR in the high-field region may arise due to the Adler-Bell-Jackiw anomaly in which LMR decreases due to the chiral imbalance \cite{WAL_crossover,ABJ}.

We have also measured LMR along the $c$ axis to understand the downward trends of MR as seen in Fig. \ref{MR_doped}(b). Both the current and the magnetic field are applied along the $c$ axis in this configuration. The field-dependent LMR at various temperatures in such a configuration is shown in Fig. \ref{MR_doped}(c). The LMR exhibits a WAL effect in the low field region, but as the field increases, it starts to decrease and becomes  negative at higher fields. Furthermore, we have measured field-dependent MR at 4 K for different angular positions of the magnetic field with respect to the \textit{c} axis when current is also along the $c$ axis as shown in Fig. \ref{MR_doped}(d). A negative MR is observed in the higher fields when $\theta$ = 0$^\circ$ (i.e. $B\parallel I\parallel c$ axis). With the increase of $\theta$, MR becomes less negative, and at $\theta$ $\geq$ 75$^\circ$, MR becomes positive. For $\theta = 90^\circ$ (i.e. $B\perp I\parallel c$ axis), MR increases monotonically with the increase of the field. The observed downturn in LMR below 60 K [see Fig. \ref{MR_doped}(b)] may be due to the chiral anomaly effect induced from a small component of the current and magnetic field along the \textit{c} axis as a result of a small misalignment which is unavoidable. 

The negative MR may originate from the chiral anomaly effect. Our band structure calculation predicts that  CaAu$_{0.5}$Cu$_{0.5}$As hosts triply degenerate points near the Fermi level. It has been demonstrated that when the current and the magnetic field are applied along the $C_3$ rotation axis of a crystal, a triply degenerate point splits into Weyl points due to Zeeman coupling, leading to the chiral anomaly effect \cite{Zhu,Chang2017,Ma2018}. So, the chiral anomaly-induced negative LMR is expected in the case of CaAu$_{0.5}$Cu$_{0.5}$As as the $C_3$ rotation axis is along the $c$ axis. Recently, the chiral anomaly-induced negative MR was observed in triply degenerate semimetals like WC \cite{WC} and YRh$_6$Ge$_4$ \cite{YRh6Ge4}. Note that the other possible source of negative LMR is the current jetting effect. To rule out such a possibility, the Ohmic contacts were made on the surface of the samples across a line using sliver paint to ensure homogeneous current distribution. Furthermore, several independent measurements were performed on the same and different crystals, and the data were reproducible within the experimental error.
\begin{figure}
	\includegraphics[width=8.5cm, keepaspectratio]{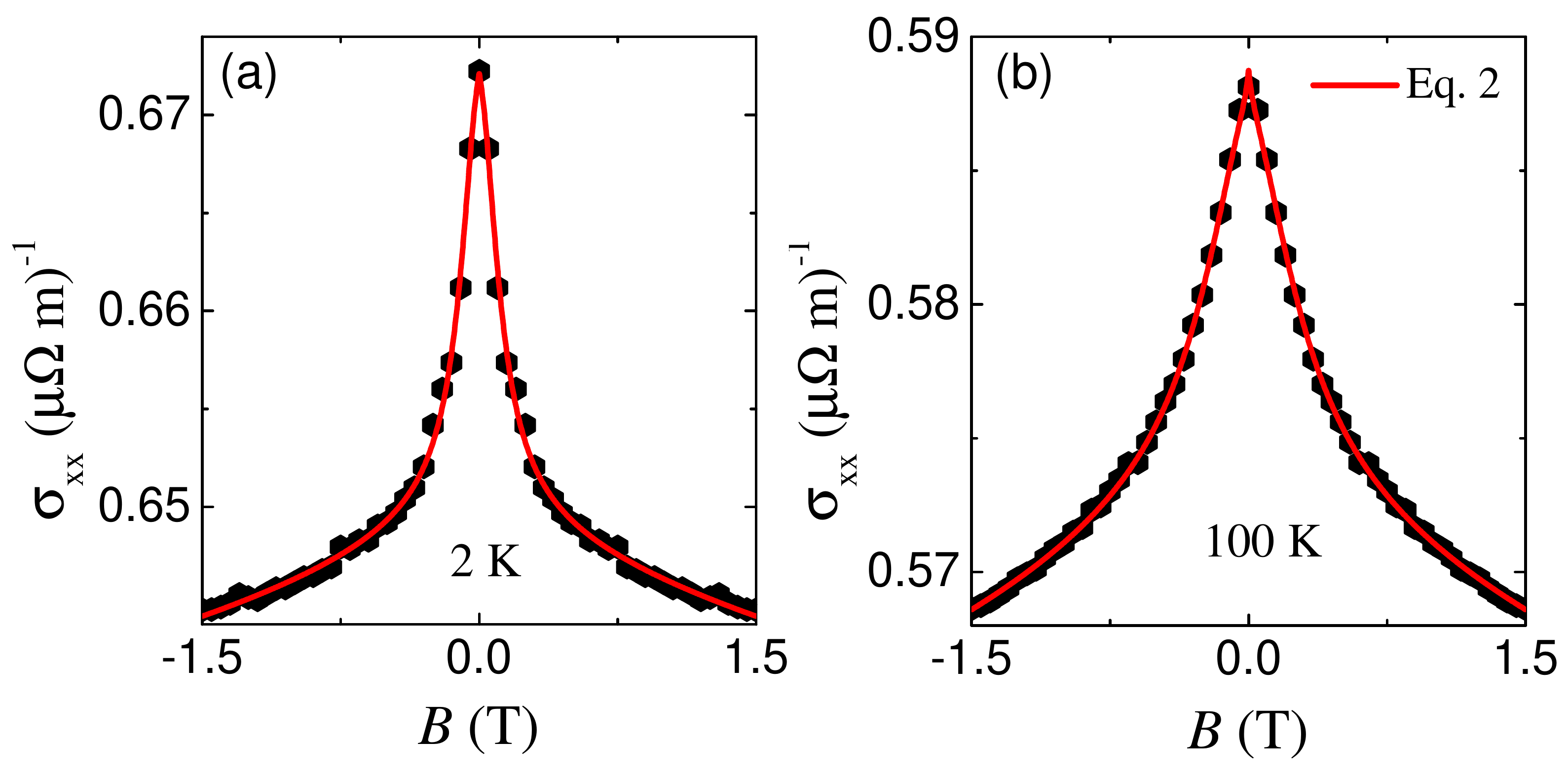}
	\caption{\label{SigmaT} MC data of CaAu$_{0.5}$Cu$_{0.5}$As measured at (a) 2 K and (b) 100 K. The red line is the fit to the 3D WAL model (Eq. \ref{Eq2}).}
\end{figure}

To analyze the WAL effect, we have calculated magnetoconductivity (MC) from the TMR data [Fig. \ref{MR_doped}(a)] using the relation $\sigma_{xx}(B) = \rho_{xx}/(\rho_{xx}^2+\rho_{yx}^2)$, where $\rho_{xx}$ and  $\rho_{yx}$ are the longitudinal and Hall resistivities, respectively. According to the semiclassical model, the transverse MC in the presence of the WAL effect for a 3D system can be expressed as \cite{3D_WAL}
\begin{equation}
\label{Eq2}
\sigma_{T}(B) = \sigma_{WAL} + \sigma_{n},
\end{equation}

where $\sigma_{WAL}$ is the contribution to conductivity due to WAL, which is expressed as $\sigma_{WAL}$=$\sigma_{xx}(0) + \eta\sqrt{B}$, and $\sigma_{n}=[\rho_{xx}(0) + A\cdot B^2]^{-1}$ is the correction to conductivity that comes from the conventional Fermi surface. $\sigma_{xx}(0)$, $\rho_{xx}(0)$, $\eta$ and $A$ are the zero-field conductivity, resistivity, and constants, respectively. An excellent agreement of MC data with Eq. \ref{Eq2} in the field range -1.5 T $\leq$ $B$ $\leq$ 1.5 T, confirms the presence of WAL effect in CaAu$_{0.5}$Cu$_{0.5}$As. As a representative,  two fitted curves of MC data calculated from TMR are presented in Fig. \ref{SigmaT} and the values of fitting parameters are $\sigma_{xx}(0)$ = 0.655 $(\mu\Omega$ m)$^{-1}$, $\rho_{xx}(0)$ = 54 $\mu\Omega$ m, $\eta$=-7.83$\times$10$^{-3}$ $(\mu\Omega$ m)$^{-1}$T$^{-0.5}$, and $A$ = 4004 $\mu\Omega$ mT$^{-2}$ at 2 K and $\sigma_{xx}(0)$ = 0.580 $(\mu\Omega$ m)$^{-1}$, $\rho_{xx}(0)$ = 106 $\mu\Omega$ m, $\eta$= -9.36 $\times$10$^{-3}$ $(\mu\Omega$ m)$^{-1}$T$^{-0.5}$, and $A$ = 1222 $\mu\Omega$ mT$^{-2}$ at 100 K, respectively.

\begin{figure}[t]
	\includegraphics[width=8.6cm, keepaspectratio]{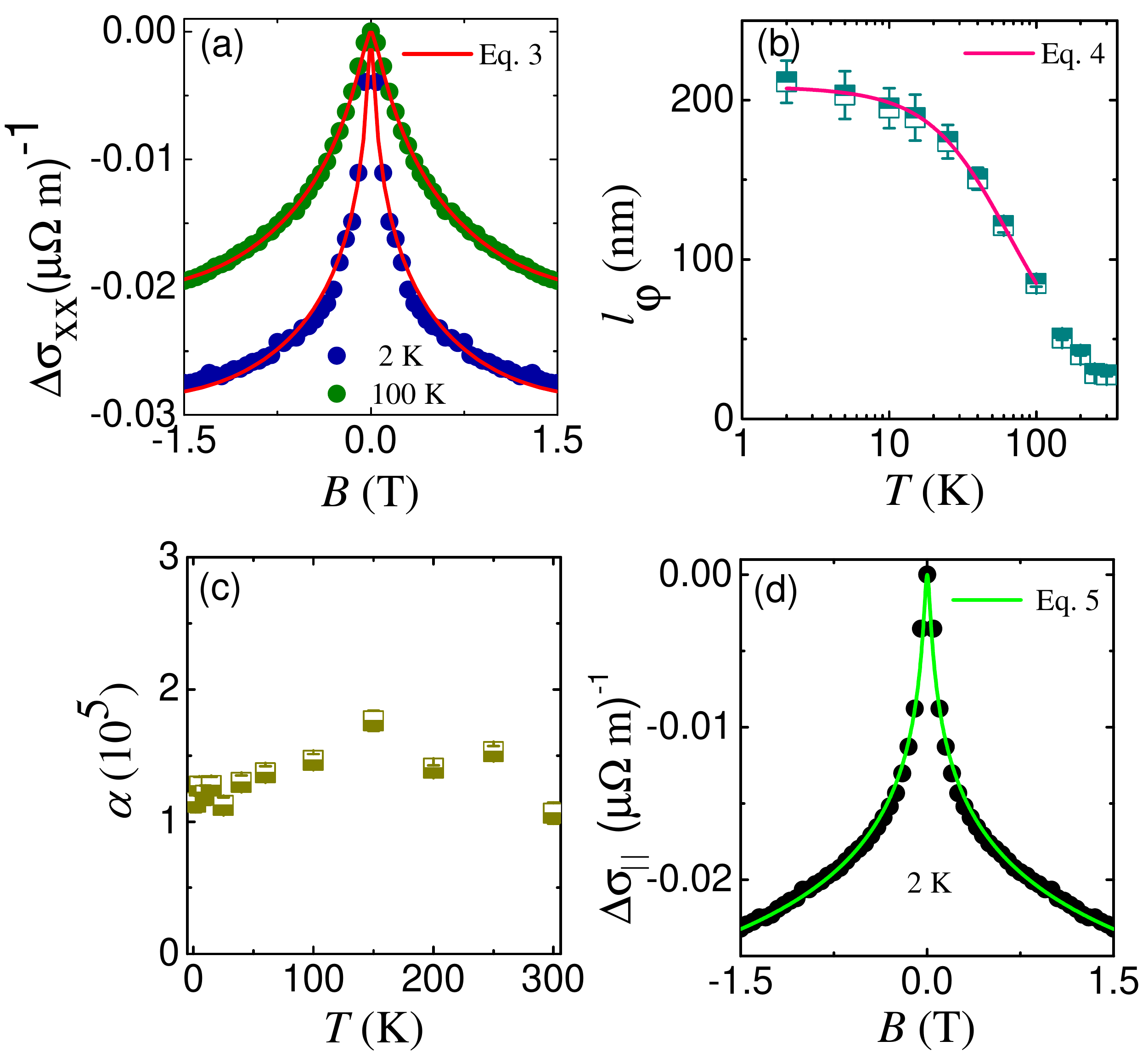}
	\caption{\label{HNL} (a) MC of CaAu$_{0.5}$Cu$_{0.5}$As is presented in the low field region at 2 and 100 K. The red line is the calculated value from the modified HNL equation. (b) The variation of $l_{\phi}$ with temperature and the pink line is the fit to Eq. \ref{Eq4}. (c) The dependence of parameter $\alpha$ with temperature. (d) Parallel field MC fitted with the generalized AA model (green line).}
\end{figure}

Further, we have analyzed the WAL effect as observed in TMR with the modified Hikami-Larkin-Nagaoka (HLN) model. Primarily HLN model explains the  WAL effect in two-dimensional (2D) system, however, it is also noticed that the WAL effect in several topological materials like YbCdGe, $R$PtBi ($R$ = rare earth), CaAgBi and LaCuSb$_2$ is in concurrence with the HLN model \cite{YbCdGe,RPtBi,CaAgBi,LaCuSb2}. In the modified HLN model, a $B^2$ term is introduced to take care of additional scattering such as spin-orbit scattering and elastic scattering. The modified HLN equation is given as \cite{HNL,Hikami_1980}

\begin{equation}
	\Delta \sigma_{xx}(B)=-C\left[ \Psi\left( \dfrac{1}{2}+\dfrac{\hbar}{4el^2_{\phi}B}\right) -ln\left(\dfrac{\hbar}{4el^2_{\phi}B}\right)\right] + \gamma B^2,
	\label{eq3}
\end{equation}
where, $C= \frac{\alpha e^2}{\pi h}$ and $\alpha$ is 1/2 in the case of 2D materials. Here, $\Psi$, $l_{\phi}$ and $\gamma$ are defined as the digamma function, phase coherent length and coefficient of the $B^2$ term, respectively. A good fitting of MC data [$\Delta \sigma_{xx} (B) = \sigma_{xx} (B) - \sigma_{xx}(0)$] with the modified HLN equation for 2 and 100 K is shown in Fig. \ref{HNL}(a) in the field range $\pm$1.5 T. This confirms the validation of the HLN model in CaAu$_{0.5}$Cu$_{0.5}$As. The estimated values of $l_{\phi}$ are displayed in Fig. 7(b) as a function of temperature. $l_{\phi}$ decreases with the increase of temperature because the inelastic scattering increases, which destroys the phase coherence. Thus, the WAL effect weakens with increasing temperature, which is evident from Fig. \ref{MR_doped}(a). The value of $l_{\phi}$ for a disordered metal is usually in the range of a few hundred nanometers as observed in CaAu$_{0.5}$Cu$_{0.5}$As. Similar values of $l_{\phi}$ were also found in other topological materials like Bi$_2$Te$_3$ \cite{Bi2Te3}, BiSbTeSe$_2$ \cite{BiSbTeSe2} and LuPtSb \cite{LuPtSb}. It may be mentioned that the value of $l_{\phi}$ is much smaller than the crystal's thickness, which implies the 3D nature of the WAL effect. The temperature dependence of $l_{\phi}$ can be interpreted by considering both electron-electron ($e-e$) and electron-phonon ($e-ph$) scattering, which follows the relation \cite{l_phi}

\begin{equation}
	\frac{1}{l_\phi^{2}(T)}=\frac{1}{l_\phi^{2}(0)}+A_{e-e}T+A_{e-ph}T^{2},
	\label{Eq4}
\end{equation}

where, $l_\phi(0)$ is the phase coherence length at 0 K, $A_{e-e}$ and $A_{e-ph}$ are the coefficient of $e-e$ and $e-ph$ scattering, respectively. According to Fig. \ref{HNL}(b), the estimated value of $l_{\phi}$ in the low-temperature region is consistent with Eq. \ref{Eq4}, when the parameters are $l_\phi(0)$ = 209 nm,  $A_{e-e}$ = 1.48 $\times$10$^{-7}$ (nm$^2$ K)$^{-1}$ and $A_{e-ph}$ = 1.01 $\times$10$^{-8}$ (nm K)$^{-2}$. Fig. \ref{HNL} (c) discloses the large value of \textcolor{blue}{$\alpha$ ($\sim$10$^5$)} calculated from the Eq. \ref{eq3}, which indicates that WAL effect mainly originates from the 3D bulk state with multiple conduction channels. Similar behavior has also been observed in other 3D systems such as LuPtSb, LuPdBi and ScPdBi \cite{LuPtSb,LuPdBi,ScPdBi}. On the other hand, it is predicted that CaAu$_{0.5}$Cu$_{0.5}$As hosts a triply degenerate topological state. Interestingly,  the WAL effect was also recently observed in another triply degenerate nodal semimetal YRh$_6$Ge$_4$ \cite{YRh6Ge4}. We believe that the WAL effect in CaAu$_{0.5}$Cu$_{0.5}$As is originates from a strong spin-orbit coupled nontrivial topological state like in Dirac/ Weyl semimetals.

\begin{figure}
	\includegraphics[width=8.6cm, keepaspectratio]{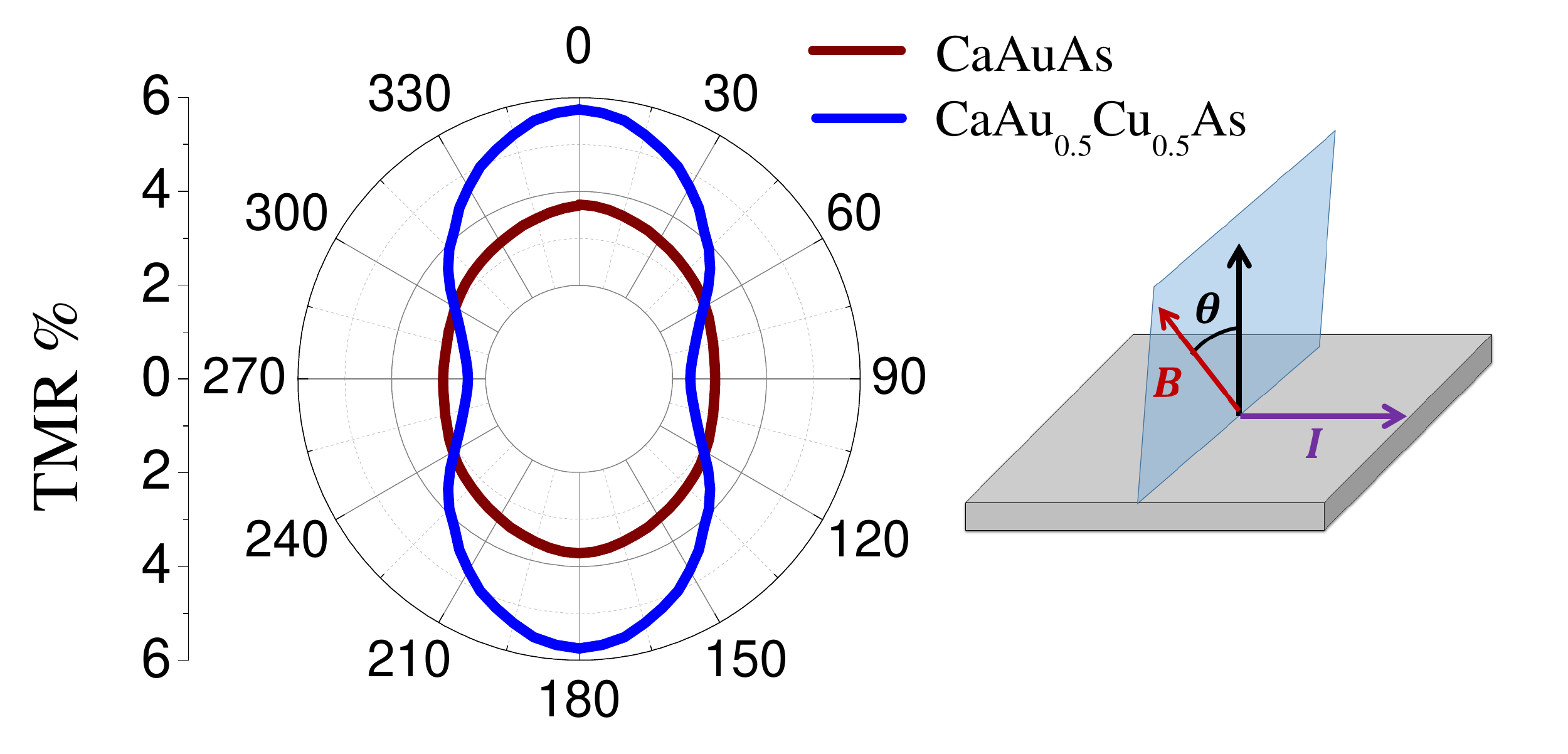}
	\caption{\label{Polar} The angular dependence of TMR of CaAuAs and CaAu$_{0.5}$Cu$_{0.5}$As measured at 5 K and 9 T are shown. The schematic diagram of experimental configuration is illustrated in the right panel.}
\end{figure}

In order to address the low-field characteristic of LMR, we analyze the parallel field MC ($\sigma_{\parallel}$) data [Fig. \ref{MR_doped}(b)] with the help of the generalized Altshuler–Aronov (AA) model. According to this model, the nature of the low-field longitudinal MC data can be described by the expression \cite{PhysRevB.88.041307}.

\begin{equation}
	\label{Eq5}
	\Delta \sigma_{\parallel} \simeq -\frac{\alpha e^2}{2\pi^2\hbar} ~ln\left[ 1+\beta\left( \frac{e^2d^2l_e^2}{\hbar^2}\right)B^2 \right],
\end{equation}

where $\Delta \sigma_{\parallel} = \sigma_{xx}(B_{\parallel}) - \sigma_{xx}(0)$. Here, $d$ (= 0.19 mm) is the sample thickness, and $l_e$ is the electron mean free path. Depending on the value of parameter $\beta$, parallel field transport can be classified into several scattering regimes. For instance, in a disordered system with $d \gg l_e$ and $\beta$ = 1/3, it is considered to be the Altshuler–Aronov regime \cite{AA}, whereas a clean metal with $d \ll l_e$ and $\beta$ = 1/6 implies the Dugaev–Khmelnitskii regime \cite{DK}. As the magnetotransport measurement has been performed on a bulk single crystal,  it is fair to consider $d \gg l_e$ \cite{EuBiTe3}. This further gives the indication that the low-field LMR of CaAu$_{0.5}$Cu$_{0.5}$As is in the AA regime. To verify the above fact, we have fitted the experimental MC data with Eq. \ref{Eq5}, as demonstrated in Fig. \ref{HNL}(d). An excellent fitting with the AA model supports our assumption. The obtained fitting parameters are $\alpha$ = 4.2$\times$10$^4$ and $l_e$ = 0.3 nm. Thus, $d \gg l_e$ and $l_\phi \gg l_e$, which indicates that the electron transport occurs in the quantum diffusive regime, which is consistent with the WAL effect \cite{Hikami_1980,Hai,Lu_2016}.

Resistivity measurements along different crystallographic directions are used to probe the anisotropic nature of the charge conduction mechanism associated with the electronic band structure. Often, the dimensions of the grown single crystals restrict the direction dependent resistivity measurements. In such a case, studying the direction-dependent TMR, by rotating the magnetic field, can be used to shed some light on the  nature of charge conduction. We have studied the angular ($\theta$) dependence of TMR of the CaAuAs and CaAu$_{0.5}$Cu$_{0.5}$As crystals. Fig. \ref{Polar} (left panel) shows TMR measured at 5 K and 9 T as a function of $\theta$. The experimental setup is also illustrated  schematically in the right panel of Fig. \ref{Polar}. In this configuration, the direction of the current is maintained along the $x$-axis within the plane of the rectangular crystal while the magnetic field is rotated slowly about the current direction ($x$-axis), i.e., from out-of-plane (along $z$-axis; $\theta$ = 0) to in-plane direction ($y$-axis; $\theta$ = 90$^\circ$). It can be seen from Fig. \ref{Polar} that the MR reaches the maximum, when field is perpendicular to the crystal plane ($\theta$ $\approx$ 0$^\circ$) and minimum when field is along the plane ($\theta$ $\approx$ 90$^\circ$). The polar plot revels a two-fold symmetric pattern, which is prominent in the case of CaAu$_{0.5}$Cu$_{0.5}$As. The estimated anisotropy ratio in TMR of CaAu$_{0.5}$Cu$_{0.5}$As at 5 K and 9 T is about 2.4, indicating a weak anisotropy in the electronic structure.

\begin{figure}[t]
	\includegraphics[width=8.6cm, keepaspectratio]{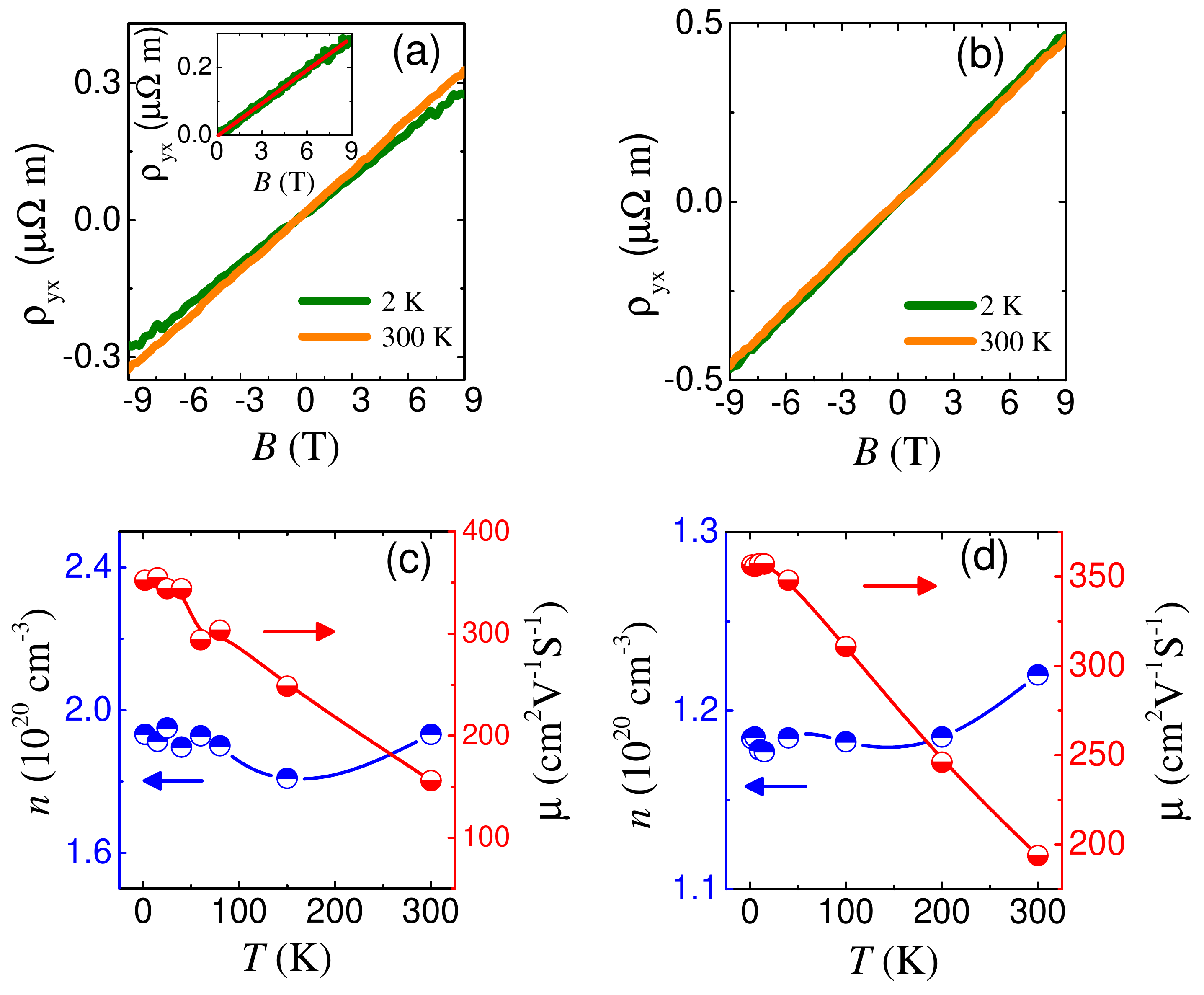}
	\caption{\label{Hall}Field dependent Hall resistivity of CaAuAs and CaAu$_{0.5}$Cu$_{0.5}$As for 2 K and 300 K is shown in (a) and (b), respectively. The inset of Fig. (a) shows the linear fit to the Hall resistivity data at 2 K. The variation of carrier concentration and mobility of both the compounds CaAuAs and CaAu$_{0.5}$Cu$_{0.5}$As are shown in (c) and (d), respectively.}
\end{figure}

To get a quantitative idea about the density of the charge carrier and its mobility, we measured Hall resistivity ($\rho_{yx}$). The field dependences of $\rho_{yx}$ for CaAuAs and  CaAu$_{0.5}$Cu$_{0.5}$As crystals at different fixed temperatures are shown in Figs. \ref{Hall}(a) and (b), respectively. Both compounds show linear Hall resistivity with a positive slope, suggesting a dominant hole type of carrier. It can also be noticed that $\rho_{yx}$($B$) is almost independent of temperature. The carrier concentration and mobility are extracted using the relations  $n = 1/eR_H$ and  $\mu= R_H/\rho_{(B=0)}$, where $R_H$ is the slope of the linear fit of $\rho_{yx}$ vs. $B$ curve [see inset of Fig. \ref{Hall}(a)]. For both compounds, the estimated values of carrier concentration and mobility as a function of temperature are shown in Figs. \ref{Hall}(c) and (d), respectively. The observed carrier concentration in both compounds is nearly independent of temperature. The estimated carrier concentration is $n \sim 10^{20}$ cm$^{-3}$, which is much smaller than in typical metals ( $n \sim 10^{22}$ cm$^{-3}$ to 10$^{23}$ cm$^{-3}$). This indicates a low density of states at the Fermi level, which further implies the semimetallic nature of both compounds as observed in similar topological semimetals like CaAgAs \cite{CaAgAs}, CaAgBi \cite{CaAgBi}, YbCdGe \cite{YbCdGe}, and PrAlGe \cite{PrAlGe}. The $\mu$ varies in between 352 cm$^2$ V$^{-1}$ S$^{-1}$ and 155 cm$^2$ V$^{-1}$ S$^{-1}$ for CaAuAs, and 355 cm$^2$ V$^{-1}$ S$^{-1}$ and 195 cm$^2$ V$^{-1}$ S$^{-1}$ for CaAu$_{0.5}$Cu$_{0.5}$As in the temperature range form 2  to 300 K. The estimated value of the mobility is a few orders of magnitude smaller than that observed in typical topological Dirac/Weyl semimetal \cite{Cd3As2,NbP}. However, the value of $\mu$ is comparable to that observed in topological materials like YbCdGe and CaAgAs.

\section{SUMMARY}

We have successfully doped 50\% Cu at the Au site of the topological Dirac semimetal CaAuAs. Interestingly,  a  significant change in the nature of MR is noticed due to Cu-doping in CaAuAs. The low-field behavior of MR of the doped system gives a clear indication of the WAL effect. MC of CaAu$_{0.5}$Cu$_{0.5}$As in the low-field regime was found to fit well with the 3D WAL model and the modified HNL model. The WAL effect in longitudinal MC was well described by the Altshuler-Aronov model. Our electronic band structure calculation suggested that the replacement of Au with 50\% Cu breaks the space inversion symmetry, which gives rise to a phase transformation from a fourfold degenerate Dirac point to two threefold degenerate triple points. The signature of a chiral anomaly further supports the existence of triple-point state in CaAu$_{0.5}$Cu$_{0.5}$As. ARPES measurements can confirm the non-trivial band topology of the alloy system.

\section{ACKNOWLEDGMENT}

We acknowledge IIT Kanpur, Science and Engineering Research Board, India (Projects No. SRG/2019/001686 and No. CRG/2018/000220), and Department of Science and Technology, India, for financial support. We thank S. Roy for the magnetoresistance measurement along the $c$ axis. ZH acknowledges the Polish National Agency for Academic Exchange for the Ulam Fellowship. A.A. acknowledges the support from the SPACE-TIME supercomputing facility hosted at IIT Bombay, where the electronic structure simulations are performed.
	
\bibliography{Reference_CaAuAs}
	
\end{document}